\documentstyle[pra,aps]{revtex}
\begin{document}
\preprint{\today}
\draft
%
%
\title{Gravitation at the mesoscopic scale
\footnotemark[1]
\footnotetext[1]
{This essay received 
``honorable mention'' from the Gravity Research Foundation, 1997 --- Ed.}
}
\author{R. Onofrio${}^1$\thanks{E-mail: Onofrio@padova.infn.it} 
and L. Viola${}^2$\thanks{E-mail:Vlorenza@mit.edu}} 
\address{${}^1$ Dipartimento di Fisica ``G. Galilei'', 
Universit\`a di Padova, and INFN, Sezione di Padova,\\  
Via Marzolo 8, Padova 35131, Italy \\
${}^2$ Department of Physics, Massachusetts Institute of Technology, 12-127, 
\\ 77 Massachusetts Avenue, Cambridge, MA 02139, USA}
\date{\today}
\maketitle
%
%
\begin{abstract}
Free fall experiments are discussed by using test masses associated to 
quantum states not necessarily possessing a classical counterpart. 
The times of flight of the Galileian experiments using classical test 
masses are replaced in the quantum case by probability distributions 
which, although still not defined in an uncontroversial manner, 
become  manifestly dependent upon the mass and 
the initial state. Such a dependence is also expected in non inertial 
frames of reference if the weak equivalence principle still holds. 
This last could be tested, merging recent achievements in mesoscopic 
physics, by using cooled atoms in free fall and accelerated frames 
initially prepared in nonclassical quantum states. 
\end{abstract}
%
%
\vspace{2mm}

\vspace{4mm}

Quantum mechanics and general relativity are still the milestones of modern  
physics in the closing part of this century. 
Their fields of applicability were not originally overlapping, since 
quantum mechanics was developed for microscopic phenomena where gravitational 
effects were expected to play a negligible role and, vice versa, gravity 
manifests in the macroscopic world where no quantum phenomena were evidenced. 
The situation has been changing in more recent times due both to 
possible astrophysical observations and controllable laboratory experiments.
Pioneered by Hawking, quantum phenomena were predicted for highly 
massive relativistic objects such as black holes \cite{HAWKING}. 
On the other hand, thanks to the astonishing development of new 
experimental techniques, the frontier of applicability of quantum mechanics 
itself has been pushed further in the macroscopic realm by crossing an 
intermediate {\sl mesoscopic} region where, albeit no more on a properly 
microscopic scale, quantum features still dominate.  
These developments, although motivated by the need to understand 
more neatly the macroscopic-microscopic borderline in quantum 
phenomena, affect gravitation as well. 
Since the current belief is that quantum laws are the fundamental 
ones, the fact that gravitational laws have been formulated in terms of  
macroscopic test bodies is accidental.  Hence it should be relevant to 
discuss how to describe gravitation operationally without mention of 
classical concepts.
This is also motivated by many attempts and proposals to investigate 
gravitation using microscopic and mesoscopic systems, such as antimatter 
in free fall \cite{FAIRBANK,NIETO}, superfluid and superconducting 
systems \cite{ANANDAN}, cooled atoms in optical molasses \cite{KASEVICH1} 
and  opto-gravitational cavities \cite{AMINOFF}, interfering matter waves  
both in curved space-time and accelerated reference frames 
\cite{COLELLA,MASHHOON,KASEVICH2,AUDRETSCH}, neutrinos acquiring 
phases of gravitational origin \cite{AHLUWALIA1}.

Our essay deals with the issue of establishing the foundational 
basis of gravitation by using quantum objects not necessarily 
possessing a classical counterpart \cite{PENROSE,VIOLA,AHLUWALIA2}. 
It turns out that, together to sharpen some well-established hypotheses and 
to show the emergence of some conceptual problems, the way is open to test 
gravitation in a Lilliput world made by atoms prepared in mesoscopic states.
 Among various aspects, we will focus on  the weak equivalence principle. 
Galileo discussed the universality of the ratio between gravitational 
and inertial masses by imagining test bodies in free fall from 
the tower of Pisa \cite{GALILEO}; since then several actual tests 
have been performed with very sensitive 
schemes \cite{WILL}, but only involving classical test bodies.
It is, therefore, natural to revisit the Pisa Gedankenexperiment by 
using properly prepared quantum objects.  
 
Given two particles with inertial masses $m_i^{(1)}$, $m_i^{(2)}$, 
$m_i^{(1)} \neq m_i^{(2)}$, we want to prepare them in such a way that any 
possible difference during the free fall motion must be ascribed to the 
effect of gravity. This is more subtle than expected due to the 
Heisenberg uncertainty principle \cite{VIOLA,EPSTEIN}. By denoting with 
$|\psi_1 \rangle$, $|\psi_2 \rangle$ the initial state for particles 1, 2 in 
the Schr\"odinger picture, we rephrase the Galileian prescription for 
initial positions and velocities by constraining the expectation values 
of the position and momentum operators in the following way:
\begin{equation}
\langle \hat{z} \rangle_{\psi_1} = \langle \hat{z} \rangle_{\psi_2} \:,
\hspace{1cm}
{\langle \hat{p}_z \rangle_{\psi_1} \over m_i^{(1)} } =
{\langle \hat{p}_z \rangle_{\psi_2} \over m_i^{(2)} }  \:, 
\label{GALILEO}
\end{equation}
where for simplicity we limit the attention to a one-dimensional 
nonrelativistic motion along the $z$-direction.  
Eq. (\ref{GALILEO}) implies, in general, 
that two different initial states $|\psi_1 \rangle \not =   
|\psi_2 \rangle$ in the single particle Hilbert space have to be selected: 
this is analogous to the classical situation, where the representative point
of the classical initial state in the phase space is different for the two 
masses. On the other hand, unlike the classical case, the probabilistic 
interpretation underlying quantum mechanics only allows to speak of 
{\sl mean} initial conditions like (\ref{GALILEO}), 
which are far from univocally determining the initial state of the 
two particles: rather, they give 
rise to equivalence classes of states, each one  
characterized by an average position and velocity and collapsing  
in the classical limit into a state  with both quantities sharply 
defined. Strictly speaking, this holds  
whenever quantum states admitting classical analogue are considered.
Earlier remarks by Einstein and Schr\"odinger showed that such a class of 
states by no means exhausts the totality of the admissible ones in the 
Hilbert space. Indeed, due to the superposition principle, states of 
intrinsically 
quantum nature arise even starting from states which in the classical 
limit correspond to macroscopically distinguishable ones.  
If $|\psi_n \rangle$, $n=1,\ldots, N$, denote a set of macroscopically 
distinct states of a given quantum system, any superposition  
$|\psi_0 \rangle = \sum_n c_n |\psi_n \rangle$ is also permitted,  with 
the complex coefficients $c_n$ ensuring normalization.
The nonclassical nature of this superposition state is manifest from 
its density matrix representation, 
\begin{equation}
\hat{\rho}= |\psi_0 \rangle \langle \psi_0| =
\sum_n |c_n|^2 |\psi_n \rangle \langle \psi_n| +
\sum_{n,m \not = n} c_n^{\ast} c_m |\psi_n \rangle \langle \psi_m| \:,  
\label{RHO}
\end{equation}
the off-diagonal terms being responsible for correlations of a purely 
quantum mechanical origin. 
In the classical limit, due to the action of decoherence mechanisms, 
interference effects are  lost, and the pure state description 
(\ref{RHO}) becomes identical to the statistical mixture characterized by the 
diagonal probability weights $|c_n|^2$ alone. 
For a generic observable its overall mean value in the state 
$|\psi_0\rangle$ is thought as formed, according to (\ref{RHO}), 
by two distinct terms, one surviving in the classical limit and 
another one made by the purely quantum average over non-diagonal 
matrix elements. 
A nice class of quantum states without classical counterpart 
is offered by the so-called Schr\"odinger cat states \cite{SCHRO}, 
whose simplest example is the coherent superposition of two macroscopically 
distinguishable states in the configurational space,  with wavefunction
\begin{equation}
\psi_0(z)={\cal N} \bigg\{ c_+ \exp\bigg( -{(z-z_0+\Delta)^2 \over 2 
\Delta_0^2}  \bigg)  
+ c_- \exp\bigg( -{(z-z_0-\Delta)^2 \over 2 \Delta_0^2} \bigg) \bigg\}\:, 
\label{CAT}  
\end{equation}
consisting of a properly normalized sum of two Gaussian peaks of width 
$\Delta_0$ at $z=z_0 \pm \Delta$, $\Delta>0$. In case $\Delta=0$ a standard
Gaussian wavepacket is recovered. By denoting with $\vartheta$ the 
relative phase between the complex coefficients $c_{\pm}$, 
the expectation values of position and momentum in the state (\ref{CAT}) are 
calculated obtaining:
\begin{equation}
\langle \hat{z} \rangle_{\psi_0}= 
z_0 - {\Delta \over \Delta_0} { |c_+|^2-|c_-|^2  \over
 \sqrt{\pi} \, |{\cal N}|^2 } \:,
\hspace{2cm}
\langle \hat{p}_z \rangle_{\psi_0}= 
- {2 \hbar \over \Delta_0^2} {\Delta \over \Delta_0} 
\exp(-\Delta^2/ \Delta_0^2)  { |c_+||c_-| \sin \vartheta  \over
\sqrt{\pi} \, |{\cal N}|^2 } \:.
\label{AVEP}
\end{equation}
By Fourier transforming (\ref{CAT}), it is seen that
no diagonal momentum contributions are present, leading to a purely quantum 
momentum in (\ref{AVEP}) with the form of a typical interference factor.
A vanishing value 
of $\langle \hat{p}_z \rangle_{\psi_0}$ is found if the relative phase 
$\vartheta = k \pi$, $k \in Z$, corresponding to the even (male, $c_+=c_-=1$) 
and odd (female, $c_+=- c_-=1$) combinations of definite parity cat states 
\cite{MANKO}; 
a maximum contribution is instead achieved at $\vartheta= (2 k+1)\pi/2$, 
$k \in Z$, corresponding to cat wavefunctions already introduced by Yurke 
and Stoler \cite{YURKE}. 
For the choice of two cat-like quantum particles of mass $m^{(1)}_i, 
m^{(2)}_i$, it is amazing to realize that prescriptions (\ref{GALILEO}), 
although no more supported by any correspondence arguments, can still be 
fulfilled  by suitably gauging the parameters of the wavefunction (\ref{CAT}). 
Having checked this possibility in general, we will hereafter shorten formulas 
by assuming $\langle \hat{p}_z \rangle_{\psi_0}=0$. 

Once the initial preparation of each particle is specified, let us 
consider the simplest case of a time evolution generated by the
Hamiltonian $\hat{H}=\hat{p}_z^2/2 m_i  + m_g g \hat{z}$, 
$m_g$ denoting the gravitational mass, and let us focus, in analogy with 
Galileo's procedure, on the times of flight of the quantum particles from the 
initial height $z_0$ to a ground reference level. Even in the absence of any 
statistical uncertainty, one must then face the fact that time of flight 
probability distributions, instead of well-defined values as in the classical
case, are in principle demanded to gain complete information on the free fall
dynamics.
It is perhaps surprising that no general consensus has been reached 
so far on a consistent definition of time of flight probability densities, 
even if various attempts have been made until recent times \cite{ROVELLI1}. 
Leaving aside a rigorous derivation, we can limit ourselves to 
semiclassical arguments.  One can then infer the average time of 
flight $T^{(k)}_{of}$ for the test mass $m_i^{(k)}$ by inverting the 
relation for the average position of the particle, which is available 
from Ehrenfest's theorem:
\begin{equation}
\langle z^{(k)}(t=T^{(k)}_{of} ) \rangle = 
z_0 - {1 \over 2} {m_g^{(k)} \over m_i^{(k)} } g t^2 =0 
\:, \hspace{1cm} k=1,2\:.       \label{FREEFALL}
\end{equation}
A rough estimate of the fluctuations around its mean value, taking into 
account the spreading of the state during the motion, can be given by 
  $\sigma_{T_{of}} \approx \sigma_z(T_{of})/v_z(T_{of})$,  
$\sigma^2_z(T_{of})$ and $v_z(T_{of})$ being the position variance  
and the average velocity at time $t=T_{of}$ respectively. 
The general expression for  
$\sigma^2_z(t)= \langle \hat{z}^2 \rangle_{\psi_t} -
\langle \hat{z} \rangle_{\psi_t}^2$ can be evaluated \cite{VIOLA}. In an 
intermediate regime where $\Delta \approx \Delta_0$ and nonclassical effects 
are maximally enhanced, the result can be cast in a simple form so that: 
\begin{equation}
T_{of}^{(k)} \pm \sigma^{(k)}_{T_{of}} = \sqrt{ 2 
\bigg( {m_i^{(k)} \over m_g^{(k)}} \bigg) {z_0 \over g} } \,\pm\, 
{\sqrt{2}\over 2}
\epsilon {\hbar \over \Delta_0 m^{(k)}_g g }\:, \hspace{1cm} k=1,2\:,
\label{DISTRIB}
\end{equation}
being $\epsilon$ a numerical factor equal to 1 for a Gaussian state and 
 to ${[(\mbox{e}-1) / (\mbox{e}+1)]}^{\pm 1/2}$ for a male ($+$) or a female 
($-$) cat state. It is manifest from (\ref{DISTRIB}) that, despite the 
semiclassical preparation recipe (\ref{GALILEO}), the time of flight 
distributions depends upon the different choices of masses and/or initial 
states. 
The ratio between inertial and gravitational mass contributes to the average 
time of flight, whereas the fluctuations around this value are affected 
by the {\sl gravitational} mass alone. Once  any state difference is 
removed by choosing initial states of the same kind, a mass dependence still 
survives even if, in close analogy with the classical experiment, 
the equality  ${m_i^{(1)}/ m_g^{(1)}} ={m_i^{(2)} / m_g^{(2)}}$
is established from observing identical values of the average time of flight. 
 In other words, the widespread quoted universal behavior in a 
gravitational field breaks down when quantum test bodies are included. 
However, 
this does not necessarily prevent quantum effects from being universally 
influenced by the interaction with gravity, provided one can interpret 
{\sl any} dependence on the mass parameter within a purely {\sl kinematical} 
framework.  
In our problem,  formally identical equations indeed hold for 
the motion of a quantum particle freely evolving in a non-inertial reference 
frame accelerating with $a=g$, and identical probability distributions are, 
therefore, predicted for similar time of flight experiments \cite{VIOLA}.
In general, universal classical behavior is replaced by mass 
dependent quantum observables, but such a dependence is expected - with an 
identical structure for motions occurring in a gravitational or an 
accelerated laboratory -  
in order Einstein's equivalence principle to be preserved at the quantum level.
Note that this specific feature can be deeply related to the impossibility 
of reproducing, for any quantum object, the classical concept of 
a {deterministic} trajectory. 
In the Nelson's picture of quantum mechanics \cite{NELSON}, where the 
kinematics is modelized through {stochastic} configurational trajectories, 
it is not surprising that the combination $\hbar/m$ 
ultimately appears in (\ref{DISTRIB}), {\it i.e.}  the Brownian 
diffusion coefficient accounting for the degree of stochasticity. 
This inevitable randomness, which preexists to the introduction of the 
gravitational field itself, is a source of troubles in extending the 
geometric interpretation which so beautifully underlies gravity at the 
classical level.  Indeed, the possibility of a simple identification 
between the world lines of freely falling bodies and a set of preferred 
entities with a purely 
geometric nature clearly no more holds \cite{SONEGO,ROVELLI2,VIOLA1}. 
At the same time, this difficulty has been also pointed out as the motivation 
to introduce a new definition of the equivalence principle \cite{LAMMER}.
Moreover, it has been recently evidenced that non-geometric features 
of gravity appear by considering quantum mechanical clocks based 
on flavor-oscillations \cite{AHLUWALIA2}.

>From the experimental viewpoint, no evidence  exists to date that the 
description given above actually fits the significance of the 
equivalence principle in the quantum realm. Nevertheless, the impressive 
progress attained in the manipulation of atomic states gives hope to make 
experimental investigation feasible. We only recall here that time of 
flight measurements on the free fall of a Gaussian cloud of cooled Cesium 
atoms were already performed in a semiclassical regime \cite{AMINOFF,DALIBARD} and,  
on the other hand, Schr\"odinger cats of a single trapped 
$\mbox{Be}^+$ ion have been also generated in laboratory \cite{WINELAND}. 
Merging these extraordinary accomplishments, a mesoscopic experimental test
of the equivalence principle can be envisaged, whereby a 
Gaussian or a Schr\"odinger cat state of matter at the single atom level 
is subjected to a gravitational or an accelerating field. 
 
The relevance of mesoscopic physics for exploring gravitation is not limited 
to a proper establishment of the equivalence principle. First, deviations from
the universal Newton law due to microscopic models of gravity have been 
proposed \cite{DAMOUR} and can be  investigated at a 
submillimeter lengthscale with micromechanical devices \cite{CARUGNO}. 
In addition, after the recent spectacular 
achievement of Bose-Einstein condensation \cite{BEC}, a new ultracold 
state of matter is available and can be used to probe the tiny 
energy scales associated to gravitationally bound states. 
As a general remark,  mesoscopic dynamics is extremely sensitive 
to perturbations introduced by the surrounding environment. Therefore, 
the effect 
of any interrogation on the system should be taken into account, ultimately 
implying that any operative statement about gravitational properties has 
to be given consistently with quantum measurement 
theory \cite{MQM,MQM1}. 
As emphasized in \cite{VIOLA}, the search for a unified picture in this 
framework will  not unlikely require the emergence of new concepts in 
gravitation. 
 
Although still in its infancy, mesoscopic gravitation gives promising 
directions to investigate gravity in a landscape where it is forced 
to convive with quantum mechanical laws.

\acknowledgments
We are grateful to Clair Mason for a critical reading of the manuscript.

\end{document}